\documentclass[11pt]{article}

\usepackage{amsmath}
\usepackage{amssymb}
\usepackage{amscd}

\usepackage[english]{babel}

\usepackage[raiselinks=false,plainpages=false,pdfpagelabels,naturalnames=true,colorlinks=true,citecolor=blue,urlcolor=blue,linkcolor=blue,bookmarksopen=true,pagebackref,hyperindex]{hyperref}

\allowdisplaybreaks[1]

\usepackage{latexsym}
\usepackage[emu]{cmll}

\def\Section#1{\section{#1}}
\def\Subsection#1{\subsection{#1}}

\def\Hide#1{\relax}

\DeclareSymbolFont{AMSb}{U}{msb}{m}{n}
\DeclareSymbolFontAlphabet{\mathbb}{AMSb}
\DeclareSymbolFont{symbolsC}{U}{txsyc}{m}{n}
\DeclareMathSymbol{\rJoin}{\mathrel}{symbolsC}{89}

\newcommand{\oCKb}[2]{#1\downarrow{#2}}

\def\VH{\mathbf{H}}

\def\VK{\mathbf{K}}
\def\VL{\mathbf{L}}

\def\Func#1{{\sf{#1}}}

\def\Iff{\Leftrightarrow}

\def\Max{\Func{max}}
\def\Min{\Func{min}}

\def\Nge{\mathrel{\ooalign{$\ge$\cr\hidewidth$|$\hidewidth}}}
\def\Nle{\mathrel{\ooalign{$\le$\cr\hidewidth$|$\hidewidth}}}

\newtheorem{Theorem}{Theorem}[section]
\newtheorem{Lemma}[Theorem]{Lemma}
\newtheorem{Corollary}[Theorem]{Corollary}

\newtheorem{Remark}{Remark}[section]

\def\Proof{\par \noindent{\bf Proof: }}

\def\EFQ{[\Func{EFQ}]}
\def\EFQ{[\Func{EFQ}]}

\def\Lnot{{{}^{\perp}}}

\title{Studying Algebraic Structures using Prover9 and Mace4}

\author{Rob Arthan\thanks{School of Electronic Engineering and Computer Science, Queen Mary University of London, Mile End Road, London E1 4NS, UK. {\tt rda@lemma-one.com}}\;\, \& Paulo Oliva\thanks{School of Electronic Engineering and Computer Science, Queen Mary University of London, Mile End Road, London E1 4NS, UK. {\tt p.oliva@qmul.ac.uk} (corresponding author)}}

\begin{document}
\maketitle

\begin{abstract}
In this chapter we present a case study, drawn from our
 research work, on the application of a fully automated
theorem prover together with an automatic counter-example
generator in the investigation of a class of algebraic structures.

We will see that these tools, when combined with human insight
and traditional algebraic methods, help us to explore the
problem space quickly and effectively. The counter-example
generator rapidly rules out many false conjectures, while
the theorem prover is often much more efficient than a human
being at verifying algebraic identities. 

The specific tools in our case study are Prover9 and Mace4;
the algebraic structures are generalisations of Heyting
algebras known as hoops. We will see how this approach helped
us to discover new theorems and to find new or improved proofs
of known results. We also make some suggestions for how one might deploy
these tools to supplement a more conventional approach to
teaching algebra.
%
\end{abstract}


\section{Introduction}\label{sec:introduction}

Prover9 is an automated theorem prover for first order logic
\cite{Prover9-mace4}.  The companion program Mace4 searches for finite models
of first order theories.  Tools such as these have been used to attack problems
in algebra for many years.A headline result was McCune's use of a predecessor
of Prover9 to find the first proof of the Robbins conjecture \cite{McCune97}.
It is noteworthy in connection with the
present chapter that the first proof relied on reductions of the conjecture
that were originally proved by Winker \cite{Winker92} using a mixture of human
reasoning and automated proof search.
Several people worked on human readable accounts of the machine-generated
proof.  \cite{Dahn98,Burris97}.

It is the goal of this chapter to illustrate an approach that we have found
very useful in research on algebraic structures: we develop
human readable proofs of new results by
mixing automated proof search and human analysis of the proofs found
by Prover9 and of examples found by Mace4. We propose that guided use
of such tools on known examples could also be of benefit in teaching algebra.

In Section \ref{sec:intro-Prover9-and-mace4} we illustrate the use of the tools
taking the theory of \emph{semilattices} as a simple example. In Section
\ref{sec:intro-hoops} we give an introduction to the class of algebraic
structures called \emph{hoops} and apply Prover9 and Mace4 to an investigation
of these algebraic structures at the level of a possible undergraduate project.

In Section \ref{sec-double-negation} of this chapter, we present some results
from our own research obtained with the assistance of Prover9 and Mace4.
Surprisingly, Prover9's machine-generated proofs can be tractable and useful
artefacts. Human insight allows us to extract new abstractions and further
conjectures, leading to an iterative interactive process for developing the
theory and producing a human-oriented account of the proofs.

The examples in this chapter are supported by a set of Mace4 and Prover9
scripts, available from \url{http://www.eecs.qmul.ac.uk/~pbo/PTMRT/}. There the
reader can also find information about obtaining Prover9 and Mace4 and
instructions for running the scripts. The examples are organised into two
separate folders, one for Section~\ref{sec:introduction} and one for
Section~\ref{sec-double-negation}. In these sections, references to named files
refer either to the scripts or the output files produced by running them in the
corresponding folder.

As a final introductory remark, we would like to stress that we
are concerned here with applications of tools.
Applications invariably suggest desirable new
features or new tools, but we are concerned here with the development
of mathematical knowledge, either in a research or a teaching context,
by exploiting potential synergy between human capabilities and the
capabilities of tools that are available now. We believe that this
kind of synergy will be required at every stage of technological
advancement and that to learn this from practical experience
is a worthwhile achievement in its own right.
We confine our thoughts on possible future developments to the tools to our
concluding remarks.

\subsection{Using Prover9 and Mace4}\label{sec:intro-Prover9-and-mace4}

In this section we will introduce Prover9 and Mace4 using the theory of semilattices as an example.

Recall that a  semilattice can be defined as a set equipped with
a least upper bound operation on pairs of elements:
i.e., a structure $(S, \cup)$ where the binary operation
$\cup$ is  associative, commutative and
idempotent, i.e., it satisfies:
\begin{align*}
x \cup (y \cup z) &= (x \cup y) \cup z \\
x \cup y &= y \cup x \\
x \cup x &= x.
\end{align*}
Prover9 and Mace4 use a common syntax for first-order logic.
In this syntax, we can formalise the semilattice axioms as
follows (see file \texttt{semilattice.ax}):
\begin{verbatim}
op(500, infix, "cup").
formulas(assumptions).
   (x cup y) cup z = x cup (y cup z).
   x cup y = y cup x.
   x cup x = x.
end_of_list.
\end{verbatim}
Here the first line declares that we will use \texttt{cup} as an
infix operator. This is followed by the axioms of our theory
(referred to as ``assumptions'' by Prover9 and its documentation)
using \texttt{cup} in place of $\cup$\footnote{Prover9 input uses only ASCII characters.}.
By convention, the letters \texttt{x}, \texttt{y} and \texttt{z} denote variables that are implicitly universally quantified.

We can now ask Prover9 to prove theorems based on these assumptions.
As a very simple example we can formulate the following goal\footnote{%
See goal script \texttt{sl-pr1.gl}.}:
\begin{verbatim}
formulas(goals).
   x cup (x cup x) = x.
end_of_list.
\end{verbatim}
If we now execute the following command:
\begin{verbatim}
   Prover9 -f  semilattice.ax sl-pr1.gl
\end{verbatim}
then Prover9 will almost instantaneously respond with the pleasing
message ``\texttt{THEOREM PROVED}''. Its output also includes
a detailed linear representation of the proof\footnote{%
See output file \texttt{sl-pr1.txt}.}:
\begin{verbatim}
1 x cup (x cup x) = x # label(non_clause) # label(goal).  [goal].
2 x cup x = x.  [assumption].
3 c1 cup (c1 cup c1) != c1.  [deny(1)].
4 $F.  [copy(5),rewrite([4(4),4(3)]),xx(a)].
\end{verbatim}
At first glance this looks daunting, but with a little work it
is possible to gain insight from it.
The Prover9 proofs
are always presented as proofs by contradiction: after stating
the goal and the assumptions that will be used to prove it,
Prover9 denies the goal by asserting the existing of a constant \texttt{c1}
that does not satisfy the equation we are trying to prove
(\texttt{!=} is the Prover9 syntax for $\neq$).
In general, the line that denies the goal will be followed by a sequence of
steps each comprising a formula that can be deduced by applying
logical inference rules to formulas given in earlier steps.
In this case, there is just one step in this sequence, in which
some rewriting has arrived at the desired contradiction \verb|$F|
(Prover9 syntax for falsehood).
In each inference step, the formula
is followed by a justification giving a precise description of
the inference rules and how they have been applied.
There is a program \texttt{prooftrans} supplied with Prover9 that can be
used to perform some useful transformations on proofs (we have
already used it in generating the proof as shown above to renumber
the steps in consecutive order). In this case, we can ask for the
rewriting steps to be expanded giving:
\begin{verbatim}
4A c1 cup c1 != c1.  [para(2(a,1),3(a,1,2))].
4B c1 != c1.  [para(2(a,1),4A(a,1))].
4 $F.  [copy(4B),xx(a)].
\end{verbatim}
\noindent It then quickly finds that repeated rewriting with the
idempotency law (the assumption, i.e., axiom stated in step 2)
gives the false (\verb|$F|) conclusion that
\texttt{c1} is not equal to itself.
The justifications in the Prover9 proofs look complicated
because they encode a very detailed description of how each inference
rule has been applied. In trying to understand the proof, one can
generally ignore most of this detail. The rule names
\texttt{copy}, \texttt{para}%
\footnote{\texttt{para} stands for ``paramodulation'', an inference rule that performs a form of equational reasoning generalising the usual notion of
using an equation to rewrite a term within formula.}%
 etc. are followed by a list of parameters
in brackets identfying the steps that provided the inputs (antecedents)
to the rule the list of letters and numbers in brackets that come
immediately after the step number identify exactly how the rule
was applied to the formula of that step, but in most cases that
is obvious and this information can be ignored.

In a semilattice, one defines a relation $\ge$ by
\[
 x \ge y \quad \Iff \quad x \cup y = x.
\]
We can formalise this definition in Prover9 syntax
as follows (see file \texttt{sl-ge-def.ax}):
\begin{verbatim}
formulas(assumptions).
   x >= y <-> x cup y = x.
end_of_list.
\end{verbatim}
We can now get Prover9 to prove some more interesting properties.
E.g. the transitivity of $\ge$\footnote{%
See goal script \texttt{sl-trans.gl}.}:
\begin{verbatim}
formulas(goals).
   x >= y & y >= z -> x >= z.
end_of_list.
\end{verbatim}
If we execute the command:
\begin{verbatim}
   Prover9 -f  semilattice.ax sl-ge-def.ax sl-trans.gl
\end{verbatim}
then Prover9 will again quickly respond with ``\texttt{THEOREM PROVED}''.
In this case the proof it finds has 20 steps\footnote{%
See output file \texttt{sl-trans.txt}.}. As always it is a proof
by contradiction, but with a little analysis it is easy to extract
the elements of a human-readable direct proof from it.
As we will see later in this chapter, it is not always easy to
extract human-readable proofs from the Prover9 output, but analysis
of the machine-generated proofs often leads to useful insights.

The reader may well ask what happens if we ask Prover9 to prove
a false conjecture. What if we were to conjecture that the ordering
relation on a semilattice is a total order? We would express this
goal in Prover9 as follows (see file \texttt{sl-total.gl}):
\begin{verbatim}
formulas(goals).
   x >= y | y >= x.
end_of_list.
\end{verbatim}
Here ``\verb"|"'' denotes disjunction, and, recalling that variables are implicitly universally quantified, in textbook logical notation the goal is
$\forall x \forall y(x \ge y \lor y \ge x)$.
If we run Prover9 on this goal it will
necessarily fail to deduce a contradiction: it will either fail
to terminate or terminate without finding a proof (either because
no new formulas can be generated or because it has reached
a user-specified bound on execution time or memory usage).
In this case the companion program Mace4 does something much more useful.
If we execute the command:
\begin{verbatim}
   mace4 -p 1 -f semilattice.ax sl-ge-def.ax sl-total.gl
\end{verbatim}
then the Mace4 program will search for a finite semilattice that
provides a counter-example to our false conjecture.
It quickly finds one and in its log of the search process it
prints out the counter-example as follows:
\begin{verbatim}
 cup :
      | 0 1 2
    --+------
    0 | 0 2 2
    1 | 2 1 2
    2 | 2 2 2

 >= :
      | 0 1 2
    --+------
    0 | 1 0 0
    1 | 0 1 0
    2 | 1 1 1
\end{verbatim}
Here we see that Mace4 uses $0, 1, 2 \ldots$ to represent the
elements of the model, which in this case has size 3. The model
is the smallest example of a semilattice that is not totally
ordered. It is made up of two incomparable atoms $0$ and $1$
together with  their least upper bound $2 = 0 \cup 1$.

In conjunction with two companion programs \texttt{isofilter}
and \texttt{interpformat}, Mace4 can enumerate all the models
of a given size. The \texttt{Makefile} contains the commands
to do this for all semilattices with at most 5 elements.
This is particularly useful when investigating more complex
algebraic structures as generating examples by hand is often
error prone, particularly if associative operators are involved
as associativity is time-consuming to check.

\subsection{Investigating the algebraic structure of hoops}
\label{sec:intro-hoops}

In Section~\ref{sec:intro-Prover9-and-mace4}, we took semilattices
as our running example for reasons of simplicity. However,
semilattices are a little too simple to demonstrate the real power
of tools such as Prover9 and Mace4. In this section we introduce the algebraic
structures called hoops \cite{blok-ferreirim00,bosbach69a,BO} that will provide the running example for the rest
of the paper. Hoops are a generalisation of Heyting algebras (used in the study of intuitionistic logic\footnote{In a Heyting algebra one normally uses $x \to y$ for $y \ominus x$, and $x \wedge y$ for $x \oplus y$.}). They are of considerable interest, e.g., in connection with {\L}ukasiewicz logic and fuzzy logic, 
and there are many difficult open problems concerning them. 

These structures were originally investigated, first by Bosbach \cite{bosbach69a}, and
independently by B\"uchi and Owens \cite{BO}. In this section, we present an elementary investigation of
hoops using Prover9 and Mace4, and indicate how one might put these tools to use at the level of an undergraduate project
assignment.

A hoop\footnote{%
Strictly speaking this is a {\em bounded} hoop: an unbounded hoop omits
the constant $1$ and axiom (\ref{ax:efq}). We are only concerned with
bounded hoops in this book, so for brevity, we drop the word ``bounded''.}
 is a structure $(H, 0, 1, \oplus, \ominus)$ satisfying the following axioms:
\begin{align}
x \oplus (y \oplus z) &= (x \oplus y) \oplus z        \label{ax:assoc} \\
x \oplus y &= y \oplus x                              \label{ax:comm} \\
x \oplus 0 &= x                                       \label{ax:id} \\
x \ominus x &= 0                                      \label{ax:minus-self} \\
(x \ominus y) \ominus z &= x \ominus (y \oplus z)     \label{ax:curry} \\
x \oplus (y \ominus x) &= y \oplus (x \ominus y)      \label{ax:cwc} \\
0 \ominus x &= 0                                      \label{ax:zero-minus} \\
x \ominus 1 &= 0                                      \label{ax:efq}
\end{align}

The example scripts in the supporting material for this section contain a
Prover9 formalisation of the hoop axioms (file \texttt{hoop-eq-ax}) and various
goals for Prover9 and Mace4.

Axioms (\ref{ax:assoc}), (\ref{ax:comm}) and (\ref{ax:id}) are very familiar as
the axioms for a commutative monoid with binary operation $\oplus$ and identity
element $0$.  Axioms (\ref{ax:minus-self}) and (\ref{ax:curry}) are reminiscent
of properties of subtraction in a commutative group, but the remaining axioms
are less familiar.

Axiom (\ref{ax:cwc}) says that the operation $\cup$ defined by $x \cup y = x
\oplus (y \ominus x)$ is commutative.  Using axioms (\ref{ax:id}) and (\ref{ax:minus-self}) we see that this operation is also idempotent. We might
conjecture that $\cup$ is also associative, so that it makes any hoop into a semilattice, and Prover9 will readily prove this for us\footnote{%
See output file \texttt{hp-semilattice.txt}.}.
The proof is short but intricate.
This semilattice structure induces an ordering on a hoop which turns
out to be equivalent to defining $x \ge y$ to hold when\footnote{%
See output file \texttt{hp-ge-sl.txt}.} $y \ominus x = 0$.

Using Mace4, we can quickly generate examples of hoops.
Tables~\ref{tbl:hoops-of-order-2-and-3} and~\ref{tbl:hoops-of-order-4}
are based on the Mace4 output and show all hoops with between 2 and 4
elements. Inspection of these tables is very instructive, particularly
if one looks at the interactions between the order structure and
the algebraic operations.

\begin{table}
$$
\begin{array}{c@{\quad\quad}c@{\quad\quad}c@{\quad\quad}{c}}
\mbox{\bf Ordering} &
\multicolumn{2}{c}{\mbox{\bf Operation Tables}} &
\mbox{\bf Name}
\\
0 <  1 &
\begin{array}{c|cc}
   {\oplus}
       & 0 & 1 \\\hline
    0  & 0 & 1 \\
    1  & 1 & 1
\end{array}
&
\begin{array}{c|cccc}
   {\ominus}
           & 0 & 1 \\\hline
    0      & 0 & 0 \\
    1      & 1 & 0
\end{array}
&
\VL_2
\\\ \\
0 < a < 1 &
\begin{array}{c|cccc}
   {\oplus}
       & 0 & a & 1 \\\hline
    0  & 0 & a & 1 \\
    a  & a & 1 & 1 \\
    1  & 1 & 1 & 1
\end{array}
&
\begin{array}{c|cccc}
   {\ominus}
           & 0 & a & 1 \\\hline
    0      & 0 & 0 & 0 \\
    a      & a & 0 & 0 \\
    1      & 1 & a & 0
\end{array}
&
\VL_3
\\\ \\
0 < a < 1 &
\begin{array}{c|cccc}
   {\oplus}
       & 0 & a & 1 \\\hline
    0  & 0 & a & 1 \\
    a  & a & a & 1 \\
    1  & 1 & 1 & 1
\end{array}
&
\begin{array}{c|cccc}
   {\ominus}
           & 0 & a & 1 \\\hline
    0      & 0 & 0 & 0 \\
    a      & a & 0 & 0 \\
    1      & 1 & 1 & 0
\end{array}
&
\VL_2 \frown \VL_2
\end{array}
$$
\caption{Hoops of order 2 and 3}
\label{tbl:hoops-of-order-2-and-3}
\end{table}

\begin{table}
$$
\begin{array}{c@{\quad\quad}c@{\quad\quad}c@{\quad\quad}{c}}
\mbox{\bf Ordering} &
\multicolumn{2}{c}{\mbox{\bf Operation Tables}} &
\mbox{\bf Name}
\\
0 < a < b < 1 &
\begin{array}{c|cccc}
   {\oplus}
       & 0 & a & b & 1 \\\hline
    0  & 0 & a & b & 1 \\
    a  & a & a & b & 1 \\
    b  & b & b & 1 & 1 \\
    1  & 1 & 1 & 1 & 1
\end{array}
&
\begin{array}{c|cccc}
   {\ominus}
           & 0 & a & b & 1 \\\hline
    0      & 0 & 0 & 0 & 0 \\
    a      & a & 0 & 0 & 0 \\
    b      & b & b & 0 & 0 \\
    1      & 1 & 1 & b & 0
\end{array}
&
\VL_2 \frown \VL_3
\\\ \\
0 < a < b < 1 &
\begin{array}{c|cccc}
   {\oplus}
       & 0 & a & b & 1 \\\hline
    0  & 0 & a & b & 1 \\
    a  & a & b & 1 & 1 \\
    b  & b & 1 & 1 & 1 \\
    1  & 1 & 1 & 1 & 1
\end{array}
&
\begin{array}{c|cccc}
   {\ominus}
           & 0 & a & b & 1 \\\hline
    0      & 0 & 0 & 0 & 0 \\
    a      & a & 0 & 0 & 0 \\
    b      & b & a & 0 & 0 \\
    1      & 1 & b & a & 0
\end{array}
&
\VL_4
\\\ \\
0 < a < b < 1 &
\begin{array}{c|cccc}
   {\oplus}
       & 0 & a & b & 1 \\\hline
    0  & 0 & a & b & 1 \\
    a  & a & b & b & 1 \\
    b  & b & b & b & 1 \\
    1  & 1 & 1 & 1 & 1
\end{array}
&
\begin{array}{c|cccc}
   {\ominus}
           & 0 & a & b & 1 \\\hline
    0      & 0 & 0 & 0 & 0 \\
    a      & a & 0 & 0 & 0 \\
    b      & b & a & 0 & 0 \\
    1      & 1 & 1 & 1 & 0
\end{array}
&
\VL_3 \frown \VL_2
\\\ \\
0 < a < b < 1 &
\begin{array}{c|cccc}
   {\oplus}
       & 0 & a & b & 1 \\\hline
    0  & 0 & a & b & 1 \\
    a  & a & a & b & 1 \\
    b  & b & b & b & 1 \\
    1  & 1 & 1 & 1 & 1
\end{array}
&
\begin{array}{c|cccc}
   {\ominus}
           & 0 & a & b & 1 \\\hline
    0      & 0 & 0 & 0 & 0 \\
    a      & a & 0 & 0 & 0 \\
    b      & b & b & 0 & 0 \\
    1      & 1 & 1 & 1 & 0
\end{array}
&
\VL_2 \frown \VL_2 \frown \VL_2
\\\ \\
\begin{array}{c}
0 < a, b < 1 \\
a \Nle b \\
b \Nge a
\end{array} &
\begin{array}{c|cccc}
   {\oplus}
       & 0 & a & b & 1 \\\hline
    0  & 0 & a & b & 1 \\
    a  & a & a & 1 & 1 \\
    b  & b & 1 & b & 1 \\
    1  & 1 & 1 & 1 & 1
\end{array}
&
\begin{array}{c|cccc}
   {\ominus}
           & 0 & a & b & 1 \\\hline
    0      & 0 & 0 & 0 & 0 \\
    a      & a & 0 & a & 0 \\
    b      & b & b & 0 & 0 \\
    1      & 1 & b & a & 0
\end{array}
&
\VL_2 \times \VL_2
\\\ \\
\end{array}
$$
\caption{Hoops of order 4}
\label{tbl:hoops-of-order-4}
\end{table}

In all cases, one notes that the elements in the rows and columns of
the operation tables for $\oplus$ are in increasing order; in the
tables for $\ominus$ the elements in the rows are in decreasing order
while the elements in the columns are in increasing order.
This suggests the conjecture that $\oplus$ is monotonic in both its
operands,  while $\ominus$ is monotonic in its right operand and
anti-monotonic in its left operand. Prover9 quickly proves this
for us\footnote{%
See output file \texttt{hp-plus-mono.txt}, \texttt{hp-sub-mono-left.txt}
and \texttt{hp-sub-mono-right.txt}.}.

Axiom (\ref{ax:curry}) suggests (and inspection of the tables supports)
the conjecture that for any $x, y$ and $z$, $z \ge x \ominus y$ iff
$z \oplus y \ge x$. This property, an analogue of one of the laws for
manipulating inequalities in an ordered commutative group, is known
as the {\em residuation} property and is quickly proved
by Prover\footnote{%
See output file \texttt{hp-res.txt}.}.

A structure for the signature $(0, 1, \oplus, \ominus, \ge)$
such that $(0, \oplus, \ge)$ is an ordered commutative monoid
with least element $0$, greatest element $1$ and satisfying
the residuation axiom:
$$
z \ge x \ominus y \Iff z \oplus y \ge x
$$
is known as a (bounded) {\em pocrim}. One might conjecture
that any pocrim is a hoop. However this conjecture is false,
if we ask Mace4 to enumerate small pocrims\footnote{
See output file \texttt{pc-egs.txt}.
}, it finds 2 pocrims with 4
elements that are not hoops, as shown in Table~\ref{tbl:non-hoops}.
The residuation property is strictly weaker than axiom (\ref{ax:cwc}).
Inspection of the operation tables reveals the weakness: in a hoop,
if $x \ge y$ then $x = y \oplus (x \ominus y)$, but in a pocrim,
even when $x \ge y$, we can have $x < y \oplus (x \ominus y)$:
in the first example in Table~\ref{tbl:non-hoops}, $x \oplus y = 1$
unless one of $x$ and $y$ is $0$.
However, the axiomatisation of hoops via the pocrim axioms together
with axiom (\ref{ax:cwc}) is often more convenient and intuitive
than the purely equational axiomatisation. 

\begin{table}
$$
\begin{array}{c@{\quad\quad}c@{\quad\quad}c}
\mbox{\bf Ordering} &
\multicolumn{2}{c}{\mbox{\bf Operation Tables}}
\\
0 < a < b < 1 &
\begin{array}{c|cccc}
   {\oplus}
       & 0 & a & b & 1 \\\hline
    0  & 0 & 1 & 1 & 1 \\
    a  & a & 1 & 1 & 1 \\
    b  & b & 1 & 1 & 1 \\
    1  & 1 & 1 & 1 & 1
\end{array}
&
\begin{array}{c|cccc}
   {\ominus}
           & 0 & a & b & 1 \\\hline
    0      & 0 & 0 & 0 & 0 \\
    a      & a & 0 & 0 & 0 \\
    b      & b & a & 0 & 0 \\
    1      & 1 & a & a & 0
\end{array}
\\\ \\
0 < a < b < 1 &
\begin{array}{c|cccc}
   {\oplus}
       & 0 & a & b & 1 \\\hline
    0  & 0 & a & b & 1 \\
    a  & a & a & 1 & 1 \\
    b  & b & 1 & 1 & 1 \\
    1  & 1 & 1 & 1 & 1
\end{array}
&
\begin{array}{c|cccc}
   {\ominus}
           & 0 & a & b & 1 \\\hline
    0      & 0 & 0 & 0 & 0 \\
    a      & a & 0 & 0 & 0 \\
    b      & b & b & 0 & 0 \\
    1      & 1 & b & a & 0
\end{array}
\\\ \\

\end{array}
$$
\caption{Pocrims that are not hoops}
\label{tbl:non-hoops}
\end{table}

Inspection of 
Tables~\ref{tbl:hoops-of-order-2-and-3} and~\ref{tbl:hoops-of-order-4},
shows that for each $n$, there is a hoop of order $n$ that is linearly
ordered and generated by its least nonzero element, in the sense
that if $a$ is the least nonzero element, the other nonzero elements
are $a \oplus a$, $a \oplus a \oplus a$, etc. For any $n$, let us define\footnote{
We call these hoops $\VL_n$ in honour of {\L}ukasiewicz \cite{lukasiewicz-tarski30} whose multi-valued logics have a natural semantics with values in these hoops.
}
$$
\VL_n = (\{0, \frac{1}{n-1}, \frac{2}{n-1}, \ldots, 1\}, 0, 1, \oplus, \ominus)
$$
where $x \oplus y = \Min(x + y, 1)$ and $x \ominus y = \Max(x-y, 0)$.
Then $\VL_n$ is a linearly ordered hoop generated by its least non-zero
element $\frac{1}{n-1}$.
Copies of $\VL_n$ for various $n$ often appear in other hoops.
Inspection of the first hoop in Table~\ref{tbl:hoops-of-order-4} shows that,
the subset $\{0, b, 1\}$ comprises a subhoop isomorphic to $\VL_3$,
while the subset $\{0, a\}$ is isomorphic to $\VL_2$ viewed
as a structure for the signature $(0, \oplus, \ominus)$ (i.e., ignoring
the fact that $a \neq 1$).
This suggests a way of constructing new hoops from old: given hoops
$\VH_1$ and $\VH_2$ with underlying sets $H_1$ and $H_2$, we can
form what is called the {\em ordinal sum}, $\VH_1 \frown \VH_2$ of the two hoops
whose underlying set is the disjoint union $H_1 \sqcup (H_2 \setminus \{0\})$,
with $0$ (resp. $1$) given by $0 \in H_1$ (resp. $1 \in H_2$) and
with the operation tables defined to extend the operations of $\VH_1$ and $\VH_2$
so that for $h_1 \in H_1$ and $h_2 \in H_2$,
$h_1 \oplus h_2 = h_2$, $h_1 \ominus h_2 = 0$ and $h_2 \ominus h_1 = h_2$.
The column headed ``Name'' in
Tables~\ref{tbl:hoops-of-order-2-and-3} and~\ref{tbl:hoops-of-order-4}
gives an expression for each hoop as an ordinal sum or product of the hoops $\VL_n$.

Linearly ordered hoops are of particular importance.
We can see at a glance from
Tables~\ref{tbl:hoops-of-order-2-and-3} and~\ref{tbl:hoops-of-order-4}
that there is $1$ linearly ordered hoop of order $2$ and $2$ of order $3$
(and these are the only hoops of these orders), while there are $4$ linearly
ordered hoops of order $4$ (and just one other).
Mace4 tells us\footnote{%
See output file \texttt{hp-linear-egs.txt}.} that there are 8 linearly
ordered hoops of order $5$ (and two others\footnote{%
See output file \texttt{hp-egs.txt}.}).
This leads us to the following theorem.
The proof of this theorem proceeds by induction and Prover9 cannot
be expected to find a proof automatically, but if, informed by
the examples provided by Mace4, we set up the framework for the inductive
proof, then Prover9 will help us with the low-level details.

\begin{Theorem}
For each $n \ge 2$, there are $2^{n-2}$ isomorphism classes of linearly ordered hoops of order $n$.
\end{Theorem}
\Proof
We claim that any linearly ordered hoop of order $n$ is isomorphic
to an ordinal sum $\VL_{m_1} \frown \ldots \frown \VL_{m_k}$
for some $k$, $m_1, \ldots, m_k \ge 2$ such that $m_1 + \ldots + m_k - k + 1 = n$.
It is easy to see that two such ordinal sums
$\VL_{m_1} \frown \ldots \frown \VL_{m_k}$ and
$\VL_{n_1} \frown \ldots \frown \VL_{n_l}$ are isomorphic
iff $k = l$ and $m_i = n_i$, $1 \le i \le k$.
Moreover the sequences $\langle m_1, \ldots, m_k \rangle$ indexing these
ordinal sums are in one-to-one correspondence with the subsets of $\{1, \ldots,
n-2\}$ via:
$$
\langle m_1, \ldots, m_k \rangle \mapsto \{m_1 -1, m_1 + m_2 - 2, \ldots,
m_1 + \ldots + m_{k-1} -k + 1\},
$$
Hence there are, indeed, $2^{n-2}$ such hoops. We have only to prove the claim
which follows immediately by induction from the observation that any finite
linearly ordered hoop $\VH$ is isomorphic to $\VL_m \frown \VK$ for some $m$ and
some subhoop $\VK$ of $\VH$. This observation may be proved by considering the subhoop
generated by the least non-zero element of $\VH$  using the fact that (in any
hoop), if $x + x = x$ and $y \ge x$, then $x + y = y$, which Prover9 will
readily verify for us\footnote{%
See output file \texttt{hp-sum-lemma.txt}.}.
 \\

This theorem and its proof are a nice example of synergy between automated  methods and the traditional approach:
Mace4 provides examples that suggest a general conjecture and indicate a possible line of proof by induction.
While it is not able to automate the inductive proof\footnote{
Prover9 is a theorem-prover for finitely axiomatisable first-order theories:
it is not designed to work with something like the principle of induction that
can only be expressed either as an infinite axiom schema or as a second-order
property. The use of interactive proof assistants that can handle induction
is of potential interest in mathematics education, but is not the focus of
the present chapter.
There has been research on fully automated proof in higher-order
logic, but this is in its early days.
}, Prover9 can help us fill in tricky algebraic details.

We encourage readers interested in using tools such as Prover9 and
Mace4 in undergraduate teaching to use the example
scripts we have provided to inform the design of an undergraduate
project involving a guided investigation
of a more familiar class of algebraic structure, say Boolean algebras or
Heyting algebras using these tools.

\Section{Analysing Larger Proofs}
\label{sec-double-negation}

In the second part of this chapter we discuss, using the theory of hoops as a running example, how we have used Prover9 and Mace4 to explore new conjectures, and the methodology we used to analyse and ``understand" Prover9's machine generated proofs. The main challenge we want to focus on is in dealing with large Prover9 proofs, and how one should go about breaking these proofs into smaller, more intuitive and understandable steps. As a general methodology, we have adopted the process described in Figure \ref{methodology}.

\begin{figure}
\fbox{
\parbox{\textwidth}{
\begin{enumerate}
    \item Start with a new conjecture $\phi$
    \item Use Mace4 to check $\phi$ does not have trivial (small) counterexamples
    \item User Prover9 to search for a proof of $\phi$
    \begin{enumerate}
    	\item Once proof found, mine the proof for new ``concepts" and ``properties"
	\item Rerun the proof search taking these new concepts and properties as given
	\item Use knowledge learned, formulate new conjecture, and go back to 1.
    \end{enumerate}
\end{enumerate}
} }
\caption{Methodology}
\label{methodology}
\end{figure}

\Subsection{A homomorphism property for hoops}

As mentioned in Section \ref{sec:intro-hoops}, hoops generalise Heyting algebras. Defining the dual of an element as $x \Lnot = 1 \ominus x$, we have that in Heyting algebras the double-dual operation $x \mapsto x\Lnot\Lnot$ is a homomorphism. The conjecture $\phi$ we were working on, was whether this was also the case for hoops, i.e. do the following two homomorphism properties hold:
\begin{equation} \label{homo-minus}
(x \ominus y)\Lnot\Lnot = x\Lnot\Lnot \ominus y\Lnot\Lnot 
\end{equation}
and
\begin{equation} \label{homo-plus}
(x \oplus y)\Lnot\Lnot = x\Lnot\Lnot \oplus y\Lnot\Lnot
\end{equation}
Using Mace4 we were able to check in just a few minutes that no small (size 20 or below) counter-examples existed\footnote{See output files \texttt{conjectureNNSNNSNN.txt} and \texttt{conjecturePNNNNPNN.txt}.}. To our surprise, Prover9 found a proof of (\ref{homo-minus}) is just over 100 minutes\footnote{See output file \texttt{theoremNNSNNSNN-eq-expanded.txt}.}. This proof, however, is not as short as the ones we have seen in the previous section, involving around 177 steps. 

\Subsection{Discovering derived operations and their basic properties}
\label{sec-derived}

When faced with a long Prover9 derivation such as the one above, we tried to identify new concepts and intermediate steps in the proof that had intrinsic value, and could be understood in isolation. 
For instance, we noticed that the patterns $x\Lnot \oplus (x \ominus y)$ and $x \ominus (x \ominus y)$ appeared multiple times in the derivation. This led us to introduce new operations so that multiple steps in the proof could be understood as properties of these new operations. In total we found, apart from $x \cup y$, three further new derived operations:
\[
\begin{array}{lcl}
	x \cup y & \equiv & x \oplus (y \ominus x) \\[2mm]
	x \cap y & \equiv & x \ominus (x \ominus y) \\[2mm]
	y \setminus x & \equiv & (x \oplus y) \ominus x \\[2mm]
	\oCKb{x}{y} & \equiv & x \Lnot \oplus (x \ominus y)
\end{array}
\]
Our final choice of notation for these new operations came after we had studied their properties. The Prover9 symbols we used for these (in ASCII) are shown in Table \ref{symbol table}. When identifying these operations we also used our knowledge of the correspondence between hoops and Heyting algebras. For instance, $x \cap y$ in logical terms corresponds to $(y \to x) \to x$, which generalises double negation and in theoretical computer science is known as the \emph{continuation monad} \cite{Moggi:1989}.

So, according to step 3. (a) and (b) of our methodology, we looked first for basic properties of these new operations, or of their relation with the primitive operations. We come up  with six simple properties (listed in the following lemma) that we then added as axioms, and rerun the proof search.

\begin{Lemma} \label{ali-basic-lemma} The following hold in all hoops:
\begin{description}
	\item[$(i)$] $x \geq y \cap x$
	\item[$(ii)$] $x \geq x \setminus y$
	\item[$(iii)$] $(x \setminus y) \ominus x = 0$
	\item[$(iv)$] $x \oplus y = x \oplus (y \setminus x)$
	\item[$(v)$] $z \cap (y \ominus x) \geq (z \cap y) \ominus (z \cap x)$
	\item[$(vi)$] $x \ominus (x \cap y) = x \ominus y$
\end{description}
\end{Lemma}

%
%
Adding these lemmas cut the proof search time to just over 10 minutes\footnote{See output file \texttt{theoremNNSNNSNN-eq-basic-lemmas.txt}.}, and the number of steps to 132. This is still a resonably large proof, which would be hard to ``understand" as a whole. So we continued looking for more complex properties of these new defined operations. This time we focused on the following four steps in the proof script \texttt{theoremNNSNNSNN-eq-basic-lemmas.txt}, and noticed that these do have intrinsic value: 
\begin{verbatim}
126  (x ~ y) + (1 ~ x) = 1 ~ (x ~ (x ~ y)).
\end{verbatim}
states a duality between $\oCKb{x}{y}$ and $x \cap y$, i.e. $\oCKb{x}{y} = (x \cap y)\Lnot$.
\begin{verbatim}
132  (x ~ y) + (1 ~ x) = (y ~ x) + (1 ~ y).
\end{verbatim}
states the commutativity of $\oCKb{x}{y}$, i.e. $\oCKb{x}{y} = \oCKb{y}{x}$.
\begin{verbatim}
134  1 ~ (x ~ (x ~ y)) = 1 ~ (y ~ (y ~ x)).
\end{verbatim}
states the commutativity of $x \cap y$ under $(\cdot)\Lnot$, i.e. $(x \cap y)\Lnot = (y \cap x)\Lnot$.
\begin{verbatim}
153  1 ~ (x ~ (1 ~ (1 ~ x))) = 1.
\end{verbatim}
immediately implies that although $x \ominus x\Lnot\Lnot$ is not $0$ in general, we do have that $(x \ominus x\Lnot\Lnot)\Lnot\Lnot = 0$; and, as we will see, this is the crucial lemma in the proof of (\ref{homo-minus}).


In order to emphasise how we were able to break this long proof into a small collection of simple lemmas (each with a reasonably short proof), we will explicitly give the proof of these lemmas, and the proof of the conjecture from these lemmas. Readers who are not planning to undertake this kind of work themselves are invited to skip the details. We believe, however, that the details will be helpful to those wanting to apply a similar methodology in other contexts. \\[-2mm]

\noindent {\bf Notation}. But before we do that, let us set up a notation for naming hoop properties. We associate a letter to each of the operations as shown in Table \ref{symbol table}
\begin{table}
\begin{center}
\begin{tabular}{|c|c|c|c|}
\hline
Operator & Prover9 & Letter & Intuition  \\[1mm]
\hline
$0$ & \texttt{0} & Z  & (\emph{zero}) \\[1mm]
$1$ & \texttt{1} & O  & (\emph{one}) \\[1mm]
$\oplus$ & \texttt{+} & P  & (\emph{plus}) \\[1mm]
$\ominus$ & \verb|~| & S  & (\emph{subtraction}) \\[1mm]
$\cup$ & \texttt{cup} & J  & (\emph{join}) \\[1mm]
$\cap$ & \texttt{cap} & M  & (\emph{meet}) \\[1mm]
$\setminus$ & $\mathtt{\backslash}$ & D  & (\emph{difference}) \\[1mm]
$\oCKb{}{}$ & \texttt{nand}& A  & (\emph{ampheck})\footnotemark \\[1mm]
$(\cdot)^\Lnot$ & \texttt{(.)'} & N  & (\emph{negation}) \\
\hline
\end{tabular}
\end{center}
\caption{Nomenclature}
\label{symbol table}
\end{table}
\footnotetext{``Ampheck" from a Greek word meaning ``cutting both ways" was the name coined by C.S. Peirce for the logical NAND operation.}
and name each property by reading all the operations on the statement of the property from left to right. For instance, the property $\oCKb{x}{y} = \oCKb{y}{x}$ is named AA. Although there is a risk that two different properties will end up with the same name, this is not the case for the properties we consider here.

\Subsection{Discovering basic properties}

The first set of basic properties we discovered relate to commutativity. One of the hoop axioms states that $x \cup y$ is commutative. It is easy to construct a model, however, which shows that $x \cap y$ is not commutative in general. When analysing proofs generated by Prover9 we spotted two other interesting commutativity properties, alluded to above. The first (which we call AA as discussed above) is that $\oCKb{x}{y}$ also satisfies the commutativity property:

\begin{Lemma}[AA] \label{AA} $\oCKb{x}{y} = \oCKb{y}{x}$
\end{Lemma}
\Proof By symmetry it is enough to prove $\oCKb{x}{y} \geq \oCKb{y}{x}$. Note that $((y \ominus x) \ominus x\Lnot) = 0$, hence:
\begin{align*}
(x \ominus y) \oplus x\Lnot & = (x \ominus y) \oplus x\Lnot \oplus ((y \ominus x) \ominus x\Lnot) \tag*{Axiom (\ref{ax:id})} \\[1mm]
	& = (x \ominus y) \oplus (y \ominus x) \oplus (x\Lnot \ominus (y \ominus x)) \tag*{Axiom (\ref{ax:cwc})} \\[1mm]
	& = (x \ominus y) \oplus (y \ominus x) \oplus ((y \ominus x) \oplus x)\Lnot \tag*{Axiom (\ref{ax:curry})} \\[1mm]
	& = (x \ominus y) \oplus (y \ominus x) \oplus (y \oplus (x \ominus y))\Lnot \tag*{Axiom (\ref{ax:cwc})} \\[1mm]
	& = (y \ominus x) \oplus (x \ominus y) \oplus (y\Lnot \ominus (x \ominus y)) \tag*{Axiom (\ref{ax:curry})} \\[1mm]
	& = (y \ominus x) \oplus y\Lnot \oplus ((x \ominus y) \ominus y\Lnot) \tag*{Axiom (\ref{ax:cwc})} \\[1mm]
	& \geq (y \ominus x) \oplus y\Lnot. \tag*{Monotonicity}
\end{align*}

We also identified this interesting duality between $x \cap y$ and $\oCKb{x}{y}$:

\begin{Lemma}[MNA] \label{MNA} $(x \cap y)\Lnot = \oCKb{x}{y}$
\end{Lemma}
\Proof By Lemma \ref{ali-basic-lemma} $(i)$ we have $y \geq x \cap y$; and by $\EFQ$ we have $1 \geq x$. Hence, $(x \cap y)\Lnot \geq x \ominus y$ so that $(*) \; (x \ominus y) \ominus (x \cap y)\Lnot = 0$. Clearly we also have that $(\dagger) \; (x \ominus y) \ominus x = 0$. Therefore
\begin{align*}
(x \cap y)\Lnot 
	& = (x \cap y)\Lnot \oplus ((x \ominus y) \ominus (x \cap y)\Lnot) \tag*{$(*)$} \\[1mm]
	& = (x \ominus y) \oplus ((x \cap y)\Lnot \ominus (x \ominus y)) \tag*{Axiom (\ref{ax:cwc})} \\[1mm]
	& = (x \ominus y) \oplus ((x \ominus y) \oplus (x \cap y))\Lnot \tag*{Axiom (\ref{ax:curry})} \\[1mm]
	& = (x \ominus y) \oplus (x \oplus ((x \ominus y) \ominus x))\Lnot \tag*{Axiom (\ref{ax:cwc})} \\[1mm]
	& = (x \ominus y) \oplus x\Lnot. \tag*{$(\dagger)$}
\end{align*}

This duality when combined with Lemma \ref{AA} immediately implies that $x \cap y$ is also commutative when in the following weaker form:

\begin{Lemma}[MNMN] \label{MNMN} $(x \cap y)\Lnot = (y \cap x)\Lnot$
\end{Lemma}

Note that we make the point of giving the full proofs of all the lemmas in order to emphasise our goal of reducing the overall proof to a sequence of simple yet interesting lemmas, each of which should have reasonably short proofs (around 10 steps). The other steps in the proof \texttt{theoremNNSNNSNN-eq-basic-lemmas.txt} which we found of interest were 101, 138 and 144, which again all have short proofs as follows:


\begin{Lemma}[NPJSSO] \label{NPJSSO} $x\Lnot \oplus ((y \cup x) \ominus (y \ominus x)) = 1$
\end{Lemma}
\Proof Note that $(*) \; x\Lnot = (y \ominus x) \oplus (x\Lnot \ominus (y \ominus x))$. We have
\begin{align*}
1 & \geq x\Lnot \oplus ((y \cup x) \ominus (y \ominus x)) \tag*{Axiom (\ref{ax:efq})} \\[1mm]
	& = x\Lnot \oplus ((y \oplus (x \ominus y)) \ominus (y \ominus x)) \tag*{Def. $\cup$} \\[1mm]
	& = (y \ominus x) \oplus (x\Lnot \ominus (y \ominus x)) \oplus ((y \oplus (x \ominus y)) \ominus (y \ominus x)) \tag*{$(*)$} \\[1mm]
	& = y \oplus (x \ominus y) \oplus (x\Lnot \ominus (y \ominus x)) \oplus ((y \ominus x) \ominus (y \oplus (x \ominus y))) \tag*{Axiom (\ref{ax:cwc})} \\[1mm]
	& = x \oplus (y \ominus x) \oplus (x\Lnot \ominus (y \ominus x)) \oplus ((y \ominus x) \ominus (y \oplus (x \ominus y))) \tag*{Axiom (\ref{ax:cwc})} \\[1mm]
	& \geq x \oplus (y \ominus x) \oplus (x\Lnot \ominus (y \ominus x)) \tag*{Monotonicity} \\[1mm]
	& = x \oplus x\Lnot \oplus ((y \ominus x) \ominus x\Lnot) \tag*{Axiom (\ref{ax:cwc})} \\[1mm]
	& \geq x \oplus x\Lnot \tag*{Monotonicity} \\[1mm]
	& \geq 1. \tag*{Residuation} \\[-8mm]
\end{align*}

\begin{Lemma}[NSNSM] \label{NSNSM} $x\Lnot = (x \ominus y)\Lnot \ominus (y \cap x)$
\end{Lemma}
\Proof Note that $(*) \; (x \ominus y) \ominus x = 0$. Hence
\begin{align*}
x\Lnot 
	& = (x \oplus ((x \ominus y) \ominus x))\Lnot \tag*{$(*)$} \\[1mm]
	& = ((x \ominus y) \oplus (x \ominus (x \ominus y)))\Lnot \tag*{Axiom (\ref{ax:cwc})} \\[1mm]
	& = ((x \ominus y) \oplus (x \cap y))\Lnot \tag*{Def. $\cap$} \\[1mm]
	& = (x \cap y)\Lnot \ominus (x \ominus y) \tag*{Axiom (\ref{ax:curry})} \\[1mm]
	& = (y \cap x)\Lnot \ominus (x \ominus y) \tag*{Lemma \ref{MNMN}} \\[1mm]
	& = (x \ominus y)\Lnot \ominus (y \cap x). \tag*{Axioms (\ref{ax:comm}) and (\ref{ax:curry})}
\end{align*}

\begin{Lemma}[NNSSNN] \label{NNSSNN} $x\Lnot = x\Lnot \ominus (x \ominus x\Lnot\Lnot)$
\end{Lemma}
\Proof It is easy to show that $(*) \; x \oplus (x\Lnot \ominus (x \ominus x\Lnot\Lnot)) = 1$. Let us use the abbreviation $X = x \ominus x\Lnot\Lnot$. It is also easy to see that $(\dagger) \; ((X\Lnot \ominus x) \ominus ((x \oplus (x\Lnot \ominus X)) \ominus x)) = 0$. Hence
\begin{align*}
x\Lnot
	& = 1 \ominus x \tag*{Def. $(\cdot)\Lnot$} \\[1mm]
	& = (x \oplus (x\Lnot \ominus (x \ominus x\Lnot\Lnot))) \ominus x \tag*{$(*)$} \\[1mm]
	& = ((x \oplus (x\Lnot \ominus (x \ominus x\Lnot\Lnot))) \ominus x) \oplus ((X\Lnot \ominus x) \ominus ((x \oplus (x\Lnot \ominus X)) \ominus x)) \tag*{$(\dagger)$} \\[1mm]
	& =((x \ominus x\Lnot\Lnot)\Lnot \ominus x) \oplus (((x \oplus (x\Lnot \ominus X)) \ominus x) \ominus (X\Lnot \ominus x)) \tag*{Axiom (\ref{ax:cwc})} \\[1mm]
	& = (x \ominus x\Lnot\Lnot)\Lnot \ominus x \tag*{Axiom (\ref{ax:curry})} \\[1mm]
	& = x\Lnot \ominus (x \ominus x\Lnot\Lnot). \tag*{Axiom (\ref{ax:curry})} \\[-4mm]
\end{align*}

The above three lemmas are interesting, in the sense that it describes properties of the duality operation $x\Lnot$, either showing equivalent ways of writing $x\Lnot$, or how it relates to other complex expressions.

Finally, the crucial lemma of the proof shows that, although $x \ominus x\Lnot\Lnot \neq 0$ in general, we always have $1 \ominus (x \ominus x\Lnot\Lnot) = 1$.

\begin{Lemma}[SNNNO] \label{SNNNO} $(x \ominus x\Lnot\Lnot)\Lnot = 1$
\end{Lemma}
\Proof Note that $(*) \, x\Lnot \oplus ((x \ominus x\Lnot) \ominus x\Lnot\Lnot) = x\Lnot$ since $(x \ominus x\Lnot) \ominus x\Lnot\Lnot = 0$. Hence,
\begin{align*}
(x \ominus x\Lnot\Lnot)\Lnot
	& = (x \ominus x\Lnot\Lnot)\Lnot \oplus (x\Lnot \ominus x\Lnot) \tag*{Easy} \\[1mm]
	& = (x \ominus x\Lnot\Lnot)\Lnot \oplus ((x\Lnot \oplus ((x \ominus x\Lnot) \ominus x\Lnot\Lnot)) \ominus x\Lnot) \tag*{$(*)$} \\[1mm]
	& = (x \ominus x\Lnot\Lnot)\Lnot \oplus ((x\Lnot \oplus ((x \ominus x\Lnot\Lnot) \ominus x\Lnot)) \ominus x\Lnot) \tag*{Axioms (\ref{ax:comm}) and (\ref{ax:curry})} \\[1mm]
	& = (x \ominus x\Lnot\Lnot)\Lnot \oplus ((x\Lnot \cup (x \ominus x\Lnot\Lnot)) \ominus x\Lnot) \tag*{Def. $\cup$} \\[1mm]
	& = (x \ominus x\Lnot\Lnot)\Lnot \oplus ((x\Lnot \cup (x \ominus x\Lnot\Lnot)) \ominus x\Lnot) \tag*{Def. $\cup$} \\[1mm]
	& = (x \ominus x\Lnot\Lnot)\Lnot \oplus ((x\Lnot \cup (x \ominus x\Lnot\Lnot)) \ominus (x\Lnot \ominus (x \ominus x\Lnot\Lnot))) \tag*{Lemma \ref{NNSSNN}} \\[1mm]
	& = 1. \tag*{Lemma \ref{NPJSSO}} \\[-8mm]
\end{align*}

Lemma \ref{SNNNO} immediately implies step 159 of \texttt{theoremNNSNNSNN-eq-basic-lemmas.txt}, namely:

\begin{Lemma}[SSNNSNO] \label{SNNNO instance} $((x \ominus y) \ominus (x\Lnot\Lnot \ominus y))\Lnot = 1$
\end{Lemma}
\Proof We have
\begin{align*}
1
	& = (x \ominus x\Lnot\Lnot)\Lnot  \tag*{Lemma \ref{SNNNO}} \\[1mm]
	& \leq ((x \ominus x\Lnot\Lnot) \ominus (y \ominus x\Lnot\Lnot)) \Lnot \tag*{Monotonicity} \\[1mm]
	& = (x \ominus (x\Lnot\Lnot \oplus (y \ominus x\Lnot\Lnot))) \Lnot \tag*{Axiom (\ref{ax:curry})} \\[1mm]
	& = (x \ominus (y \oplus (x\Lnot\Lnot \ominus y))) \Lnot \tag*{Axiom (\ref{ax:cwc})} \\[1mm]
	& = ((x \ominus y) \ominus (x\Lnot\Lnot \ominus y))\Lnot \tag*{Axiom (\ref{ax:curry})} \\[1mm]
	& \leq 1 \tag*{} \\[-4mm]
\end{align*}

\Subsection{Producing a human-readable proof of (\ref{homo-minus})}

We are now in a position where we can derive a human-readable proof of the homomorphism property (\ref{homo-minus}) using the lemmas of the previous section. Our proof is based on the one in the proof script \texttt{theoremNNSNNSNN-eq-expanded.txt}. 

\begin{Theorem}[NNSNNSNN] \label{NNSNNSNN} $x\Lnot\Lnot \ominus y\Lnot\Lnot = (x \ominus y)\Lnot\Lnot$
\end{Theorem}
\Proof Since $(x\Lnot\Lnot \ominus y) \ominus (x \ominus y) = 0$ it follows that $(*) \; (x\Lnot\Lnot \ominus y) \cap (x \ominus y) = x\Lnot\Lnot \ominus y$. Hence
\begin{align*}
x\Lnot\Lnot \ominus y\Lnot\Lnot
	& = (x\Lnot\Lnot \ominus y)\Lnot\Lnot \tag*{Lemma \ref{ali-basic-lemma} $(vi)$} \\[1mm]
	& = (1 \ominus (x\Lnot\Lnot \ominus y))\Lnot \tag*{Def. $(\cdot)\Lnot$} \\[1mm]
	& = (1 \ominus ((x\Lnot\Lnot \ominus y) \cap (x \ominus y)))\Lnot \tag*{$(*)$} \\[1mm]
	& = (((x \ominus y) \ominus (x\Lnot\Lnot \ominus y))\Lnot \ominus ((x\Lnot\Lnot \ominus y) \cap (x \ominus y)))\Lnot \tag*{Lemma \ref{SNNNO instance}} \\[1mm]
	& = (x \ominus y)\Lnot\Lnot. \tag*{Lemma \ref{NSNSM}}
\end{align*}
%

\Subsection{Tackling the harder conjecture (\ref{homo-plus})}

We have also been able to produce a human-readable proof of (\ref{homo-plus}), although this seems to be a much harder result. Prover9 is also able to find a proof of (\ref{homo-plus}) from the basic hoop axioms, but that takes almost 7 hours, and requires 624 steps\footnote{See file \texttt{theoremPNNNNPNN-eq.txt}.}. In a process of ``mining" this proof for new lemmas, and then searching for a proof again using these lemmas, we were able to find a small set of lemmas from which Prover9 derives the proof of the conjecture in just a fraction of a second\footnote{See file \texttt{theoremPNNNNPNN-eq-lemmas.txt}.}. Again, in order to emphasise how we were able to produce a human-like mathematical presentation of this proof, we prove here these other lemmas in full, and give the (short) proof of (\ref{homo-plus}) from these lemmas.

The first two lemmas enable us to rewrite $x$ or $x\Lnot$ as a sum of two other elements. In a hoop we do not have idempotence ($x = x \oplus x$) in general, but we can obtain weaker forms of this as follows:

\begin{Lemma}[MPS] \label{MPS} $x = (x \cap y) \oplus (x \ominus y)$
\end{Lemma}
\Proof We have
\begin{align*}
x
	& = x \oplus ((x \ominus y) \ominus x) \tag*{Axiom (\ref{ax:id})} \\[0mm]
	& = (x \ominus (x \ominus y)) \oplus (x \ominus y) \tag*{Axiom (\ref{ax:cwc})} \\[1mm]
	& = (x \cap y) \oplus (x \ominus y). \tag*{\mbox{Def. $\cap$}}
\end{align*}

\begin{Lemma}[NSPJN] \label{NSPJN} $x\Lnot = (y \ominus x) \oplus (x \cup y)\Lnot$
\end{Lemma}
\Proof That $x\Lnot \geq (y \ominus x) \oplus (x \cup y)\Lnot$ follows directly since $(x \cup y)\Lnot = (y \ominus x)\Lnot \ominus x$. For the other direction we have:
\begin{align*}
x\Lnot 
	& = x\Lnot \oplus (y \ominus 1) \tag*{Axiom (\ref{ax:efq})} \\[1mm]
	& \geq x\Lnot \oplus ((y \ominus x) \ominus x\Lnot) \tag*{Easy} \\[1mm]
	& = (y \ominus x) \oplus (x\Lnot \ominus (y \ominus x)) \tag*{Axiom (\ref{ax:cwc})} \\[1mm]
	& = (y \ominus x) \oplus (x \oplus (y \ominus x))\Lnot \tag*{Axiom (\ref{ax:curry})} \\[1mm]
	& = (y \ominus x) \oplus (x \cup y)\Lnot. \tag*{Def. $\cup$}
\end{align*}

The following two lemmas can be seen as properties relating $x \oplus y$ with the new derived connectives $x \cup y$, $x \cap y$ and $x \setminus y$. 

\begin{Lemma}[PPMD] \label{PPMD} $x \oplus y = x \oplus (y \cap (y \setminus x))$
\end{Lemma}
\Proof Note that $(*)~(y \ominus (y \setminus x)) \ominus x = 0$. Hence
\begin{align*}
x \oplus y
	& = (y \setminus x) \oplus x \tag*{Lemma \ref{ali-basic-lemma} $(iv)$} \\[1mm]
	& = (y \setminus x) \oplus x \oplus ((y \ominus (y \setminus x)) \ominus x)  \tag*{$(*)$} \\[0mm]
	& = (y \setminus x) \oplus (y \ominus (y \setminus x)) \oplus (x \ominus (y \ominus (y \setminus x))) \tag*{Axiom (\ref{ax:cwc})} \displaybreak[1] \\[1mm]
	& = y \oplus ((y \setminus x) \ominus y) \oplus (x \ominus (y \ominus (y \setminus x))) \tag*{Axiom (\ref{ax:cwc})}  \\[0mm]
	& = y \oplus (x \ominus (y \ominus (y \setminus x))) \tag*{Lemma \ref{ali-basic-lemma} $(iii)$} \\[1mm]
	& = (y \cap (y \setminus x)) \oplus (y \ominus (y \setminus x)) \oplus (x \ominus (y \ominus (y \setminus x))) \tag*{Lemma \ref{MPS}} \\[1mm]
	& = (y \cap (y \setminus x)) \oplus x \oplus ((y \ominus (y \setminus x)) \ominus x) \tag*{Axiom (\ref{ax:cwc})} \\[0mm]
	& = x \oplus (y \cap (y \setminus x)). \tag*{Monotonicity}
\end{align*}

\begin{Lemma}[NPNPM] \label{NPNPM} $x\Lnot \oplus y = x\Lnot \oplus (y \cap x)$
\end{Lemma}
\Proof That $x\Lnot \oplus y \geq x\Lnot \oplus (y \cap x)$ follows directly from $y \geq y \cap x$. For the other direction, note that $x \geq y \setminus x\Lnot$. Hence, $y \cap x \geq y \cap (y \setminus x\Lnot)$. Therefore, the result follows directly from Lemma \ref{PPMD}.  \\

The above lemmas give us another interesting and useful duality between the two defined operations $x \cap y$ and $x \setminus y$:

\begin{Lemma}[JNND] \label{JNND} $(x \cup y)\Lnot = y\Lnot \setminus x$
\end{Lemma}
\Proof That $y\Lnot \setminus x \geq (x \cup y)\Lnot$ is easy to show. For the converse, observe that by Lemma \ref{NPNPM} (with $y$ and $x$ interchanged) it follows that $(*)$ $y\Lnot \geq (x + y\Lnot) \ominus (x \cap y)$. Hence
\begin{align*}
(x \cup y)\Lnot
	& = (x \oplus (y \ominus x))\Lnot \tag*{\mbox{Def. $\cup$}} \\[1mm]
	& = (y \oplus (x \ominus y))\Lnot \tag*{\mbox{Axiom (\ref{ax:cwc})}} \\[1mm]
	& = y\Lnot \ominus (x \ominus y) \tag*{\mbox{Axiom (\ref{ax:curry})}} \\[1mm]
	& \geq ((x \oplus y\Lnot) \ominus (x \cap y)) \ominus (x \ominus y) \tag*{$(*)$} \\[1mm]
	& = ((x \oplus y\Lnot) \ominus ((x \ominus y) \ominus x)) \ominus x \tag*{Axiom (\ref{ax:cwc})} \\[1mm]
	& = (x \oplus y\Lnot) \ominus x \tag*{Easy} \\[1mm]
	& = y\Lnot \setminus x. \tag*{Def. $\setminus$}
\end{align*}

The above, together with Lemma \ref{JNND}, immediately gives us another interesting commutativity property.

\begin{Corollary}[NDND] \label{NDND} $y\Lnot \setminus x = x\Lnot \setminus y$
\end{Corollary}

The final main lemma in the proof of (\ref{homo-plus}) is a surprising duality between $\ominus$ and $\oplus$.

\begin{Lemma}[SNNNPN] \label{SNNNPN} $(y \ominus x\Lnot )\Lnot = x\Lnot \oplus y\Lnot$
\end{Lemma}
\Proof That $x\Lnot \oplus y\Lnot \geq (y \ominus x\Lnot )\Lnot$ is easy to derive. For the converse, note that we have $x\Lnot\Lnot \ge y \ominus x\Lnot$, and hence
\[ (*) \; (x\Lnot\Lnot \oplus (y \ominus x\Lnot)\Lnot) \ominus x\Lnot\Lnot \ge (x\Lnot\Lnot \oplus x\Lnot\Lnot\Lnot) \ominus x\Lnot\Lnot = x\Lnot\Lnot\Lnot. \]
Hence, taking $x' = y \ominus x\Lnot$ and $y' = x\Lnot\Lnot$ in Lemma \ref{NSPJN}, we have the first line of the following chain
\begin{align*}
(y \ominus x\Lnot)\Lnot \tag*{Lemma \ref{NSPJN}}
	& = (x\Lnot\Lnot \ominus (y \ominus x\Lnot)) \oplus ((y \ominus x\Lnot) \cup x\Lnot\Lnot)\Lnot \\[1mm]	
	& = (x\Lnot\Lnot \ominus (y \ominus x\Lnot)) \oplus ((y \ominus x\Lnot)\Lnot \setminus x\Lnot\Lnot) \tag*{Theorem \ref{JNND}} \\[1mm]	
	& \geq (x\Lnot\Lnot \ominus (y \ominus x\Lnot)) \oplus x\Lnot\Lnot\Lnot \tag*{$(*)$} \\[1mm]
	& = (x\Lnot \oplus (y \ominus x\Lnot))\Lnot \oplus x\Lnot \tag*{Lemma \ref{ali-basic-lemma} $(vi)$} \\[1mm]
	& = (x\Lnot \cup y)\Lnot \oplus x\Lnot \tag*{Def $\cup$} \\[1mm]
	& = (x\Lnot\Lnot \setminus y) \oplus x\Lnot \tag*{Theorem \ref{JNND}} \\[1mm]
	& = (y\Lnot \setminus x\Lnot) \oplus x\Lnot \tag*{Corollary \ref{NDND} $(i)$} \\[1mm]
	& \geq x\Lnot \oplus y\Lnot. \tag*{Residuation}
\end{align*}

Lemma \ref{SNNNPN} above, immediately implies the homomorphism property for $x \oplus y$, since $(x \oplus y)\Lnot\Lnot = (y\Lnot \ominus x\Lnot\Lnot)\Lnot$.

\begin{Theorem}[PNNNNPNN] \label{PNNNNPNN} $(x \oplus y)\Lnot\Lnot = x\Lnot\Lnot \oplus y\Lnot\Lnot$
\end{Theorem}

\begin{Remark} It is interesting to observe that in the proof \texttt{theoremPNNNNPNN-eq.txt} of property (\ref{homo-plus}) one also finds some of the lemmas used in the proof of (\ref{homo-minus}), for instance, Lemmas \ref{MNA} (step 329) and \ref{NSPJN} (step 486),
%
%
but more interestingly, it also discovers Lemma \ref{SNNNPN} (step 633),
%
%
and uses that to derive a more general duality between $\ominus$ and $\oplus$ (step 683), namely 
\[ (x \ominus y)\Lnot = x\Lnot \oplus y\Lnot\Lnot \]
an interesting property, as in the absence of idempotence ($x = x \oplus x$), i.e., in a hoop that is not a Heyting algebra, it is usually hard to find non-trivial equivalences between non-sums and sums.
\end{Remark}

\Section{Concluding Remarks}

In Section \ref{sec:introduction} we have attempted to introduce the tools and methods we have been using by
examples at the level of an  undergraduate project. We hope this is of interest to educators and advocate
introduction of tools such as Prover9 and Mace4 into  mathematical curricula.

At a more advanced level, we have discussed our own research using Prover9 and
Mace4 to investigate algebraic structures.  It is possible to demonstrate the
provability of properties like duality, commutativity or homomorphism
properties by model-theoretic methods but these methods are not constructive,
whereas the methods discussed in Section \ref{sec-double-negation} construct
explicit equational proofs. 

In~\cite{deVilliers90}, de Villiers argues that proof has many purposes apart
from  verification, including explanation, systematization, intellectual
challenge, discovery and communication.  Tools such as Prover9 automate the
process of discovering a proof, but at first glance, the proofs that are
discovered seem inacessible to a human reader. We take this as an intellectual
challenge in its own right and claim that with human effort, judiciously
applied, we can ``mine'' explanative and systematic human-oriented proofs from
machine-generated ones, potentially leading to new insights into the problem
domain.

We have deliberately avoiding discussing potential developments of the
tools in the main body of this chapter. However, there are several obvious areas
for future investigation. Some automated support for refactoring
the machine-generated proofs could be very helpful. The refactoring steps
of interest would include separating out lemmas and retrofitting
derived notations. It is certainly of interest to speculate on possibilities
for fully automating extraction of human-readable proofs from machine-generated
proofs, but we view this as a hard challenge for Artificial Intelligence.


\bibliographystyle{plain}

\bibliography{references}

\end{document}